\documentclass[final,5p,times,twocolumn]{elsarticle}

\usepackage{amsmath}
\usepackage{amssymb}
\usepackage{graphicx}
\usepackage{float}
\usepackage{hyperref}
\usepackage{comment}
\usepackage[normalem]{ulem}
\usepackage{tikz}
\usepackage{tikz-feynman}
\tikzfeynmanset{compat=1.1.0}
\usetikzlibrary{decorations.markings, decorations.pathmorphing}

\tikzset{
  fermion thin/.style={
    draw=black,
    line width=0.5pt,
    postaction={
      decorate,
      decoration={markings, mark=at position 0.5 with {\arrow[scale=0.9]{stealth}}}
    }
  }
}

\graphicspath{{Article/Figures/}}

\journal{Annals of Physics}

\begin{document}

\begin{frontmatter}

\title{Quantum Collective Fluctuations in Correlated Fermion Systems}

\author[1,2]{S.S. Onuchin}
\author[3]{Ya.S. Lyakhova}
\author[1,2]{L.D. Silakov}
\author[1,4]{A.N. Rubtsov\corref{cor1}}

\affiliation[1]{organization={Russian Quantum Center},
            city={Moscow},
            postcode={121205},
            country={Russia}}

\affiliation[2]{organization={Moscow Institute of Physics and Technology},
            city={Dolgoprudny},
            postcode={141701},
            country={Russia}}

\affiliation[3]{organization={Laboratory of Superconducting Quantum Technologies of the National University of Science and Technology ``MISIS''},
            city={Moscow},
            postcode={119049},
            country={Russia}}

\affiliation[4]{organization={Lomonosov Moscow State University},
            city={Moscow},
            postcode={119991},
            country={Russia}}

\ead{ar@rqc.ru}

\begin{abstract}
Collective excitations in fermionic systems play a crucial role in determining their physical properties. An important challenge is to develop efficient theoretical approaches for describing these excitations and their coupling to fermionic degrees of freedom. In this work, we revisit the problem of quantifying the contributions of individual Matsubara modes of collective fluctuations to observable properties of correlated fermion systems within the framework of the Fluctuating Local Field (FLF) method. Whereas the auxiliary field in this method was previously considered only as static, we formulate its systematic extension termed Quantum FLF (Q-FLF) that incorporates selected Matsubara modes, thus tailoring it to description of quantum dynamical collective fluctuations. As a testbed, we apply the approach to a half-filled one-dimensional Hubbard chain and compute the Green’s function, the total energy, and the antiferromagnetic susceptibility. Our results demonstrate that the proposed scheme enables an efficient and selective characterization of the contributions of individual Matsubara modes. In particular, low Matsubara frequencies are found to have a quantitative impact on integrated observables such as total energy and antiferromagnetic susceptibility. At the same time, an accurate description of single-particle properties requires inclusion of higher-frequency modes.
\end{abstract}


\end{frontmatter}

\section{Introduction}
\label{introduction}

Physics of correlated fermions is a remarkably rich and complex field. Here we find intricate phase diagrams and a wide range of nontrivial phenomena. Prominent examples include high-$T_c$ cuprates \cite{highTc}, exhibiting pseudogap behavior and non-Fermi-liquid properties; heavy-fermion compounds \cite{Heavy}, characterized by quantum criticality; transition metal oxides \cite{transitionmetals}, displaying Mott transitions; and one-dimensional systems, where spin–charge separation occurs; etc. A substantial part of this physics originates from collective fluctuations and excitations, such as spinons and magnons, as well as from effective interactions between fermions mediated by bosonic collective modes \cite{bosoniccollective}.

Theoretical description of real systems of this type demands access to dynamical fluctuations. Among such approaches, an important class consists of methods in which collective bosonic excitations are introduced explicitly or treated effectively via two-particle correlation functions. Examples include diagrammatic techniques based on summations of RPA-type series \cite{Dukelsky2004}, Dynamical Mean-Field Theory (DMFT) and its extensions \cite{DMFTintro}, and others. In particular, the Dual Boson (DB) approach \cite{RUBTSOV20121320} provides a framework in which fermionic and bosonic collective excitations are treated on equal footing, incorporating both local and nonlocal correlations via introduction of additional (dual) degrees of freedom and diagrammatic corrections to extended DMFT (EDMFT). On the other hand, there is a number of works considering effective theories in which fermions interact with collective bosonic fields possessing nontrivial dynamics. For example, in the work \cite{doi:10.1073/pnas.2402052121} it is shown that, in models of strange metals, strongly overdamped bosonic modes play a crucial role. These modes can become spatially localized and govern transport properties, including resistivity. Such approaches generally deal with the full spectrum of bosonic excitations, at least in principle. However, the strength of theoretical approaches lies not in accounting for all possible processes indiscriminately, but in identifying the most relevant contributions. This requires development of physical intuition, which can be achieved also through the study of simple models, and preliminary estimations of the relative importance of different effects.

In recent years, a number of works have focused on analyzing the contribution of quantum fluctuations and particular bosonic interactions to observable quantities. For a number of systems it has been demonstrated that the essential physics is largely governed by low-energy modes and their coupling to fermions. For instance, in \cite{PhysRevLett.126.056403} the authors claim that Kondo physics can be extracted from the lowest Matsubara frequencies of the generalized charge susceptibility. The authors of another work \cite{10.21468/SciPostPhys.16.2.054} suggest a comprehensive analysis of bosonic modes' contribution on example of Hubbard atom, Anderson impurity model and DMFT solution of 1D Hubbard model, and agree that low-frequency physics is dominant in these systems. There are also evidences of crucial role of low-energy bosonic modes in quantum critical behavior \cite{Jiang2022Pseudogap}. Rigorous analysis of quantum impact of Matsubara modes on the susceptibility or physical observables of Hubbard model is presented in \cite{Watzenbock}.

In the present work, we revisit the problem of investigation the effects of collective fluctuations' quantum modes using the Fluctuating Local Field (FLF) method as a research tool. The FLF approach was originally introduced in \cite{Rubtsov2018FLF} on the example of classical Ising and Heisenberg model. Later, it was developed to describe fermionic systems \cite{rubtsov2020collective,JSNM-LR}. The basic idea of FLF approach is to model correlated fermion system with an ensemble of auxiliary (so-called parental) systems exposed to artificial field, which is in turn coupled to the order parameter relevant to the dominant instability. Tuning the amplitude of FLF field fluctuations, this method allows to efficiently describe collective behavior of the systems under consideration, and even develop perturbation theory for systems, which lack real small parameter \cite{ys24}. Flexibility of FLF allows to successively consider complex multi-channel regimes \cite{PhysRevB.108.035143,Linner2023PRB}, and multiple modes in one channel. In particular, in \cite{lyakhova2022fluctuatingMolecules}, this method was used to account for spatial inhomogeneities of fluctuations in 1D Hubbard model by considering several momentum modes.

Previously, the FLF approach was applied only with static (in imaginary time) auxiliary fields. We argue though that it can be extended to imaginary-time-dependent auxiliary field allowing to incorporate dynamical collective fluctuations. It would provide direct access to a microscopic description of the contributions of nonzero Matsubara modes of collective fluctuations. Previous works on this approach allows to claim, that scheme, which we term Quantum FLF (Q-FLF), is a potentially versatile theoretical tool able to controllably turn on and off separate fluctuation modes, and gives straight access to the impact of different fluctuation channels.

In this work we aim two goals. The first is the development of a general Quantum FLF approach. And the second goal is to test it on a correlated fermions description issue. As a model system, we consider small-size 1D Hubbard chains at half-filling. This choice is motivated by several reasons. First, this system features a well-defined fluctuation channel, which is especially strong for lattice sizes that are multiples of four due to the perfect nesting (Fig. \ref{fig:fig1}). Second, the small size allows us to neglect spatial inhomogeneities and focus exclusively on the role of quantum fluctuations' contributions. Furthermore, it is known that 1D Hubbard system is highly susceptible to fluctuations for any nonzero interaction strength \cite{PhysRevB.65.081105}, providing us direct access to nontrivial collective physics even at small and moderate $U$. In addition, we restrict ourselves to a minimal approximation by considering only the zeroth and the $\pm 1$ Matsubara modes. This allows us both to probe contributions of quantum fluctuations and to avoid excessive computational complexity.

The remainder of the paper is structured as follows. After the short reminder of basic FLF scheme in Section \ref{section:systemandform}, we construct its Matsubara extension in Section \ref{section:Q-FLF} and discuss the reasonings for the approximations we make further. Then we apply the developed approach to the analysis of single-particle excitations, specifically to the calculation of the Green’s function (Section \ref{section:GF}), and to global (in the sense of fermionic frequencies) quantities such as energy (Section \ref{section:energy}) and antiferromagnetic susceptibility (Section \ref{section:chi}). Our results show that including the $\pm 1$ Matsubara modes leads to a slight deterioration in the description of the Green’s function compared to the basic FLF method with zeroth Matsubara mode only. In contrast, for global quantities, an improvement is observed. This can be understood from the fact that the Green’s function is sensitive to a broad range of frequencies, and its accurate description requires contributions from many modes beyond the low-energy sector. We provide the discussion of these results in Section \ref{section:discussion}.

\section{System and Formalism}
\label{section:systemandform}
We start with a brief reminder of basic FLF approach on the example of one-dimensional periodic Hubbard chain with hopping constant $t$ and on-site Coulomb interaction constant $U$, which we use as a prototype model throughout this work. The action of this system  at half-filling regime, and in external antiferromagnetic field $\boldsymbol{h}_j = \boldsymbol{h} e^{i\pi j}$ with $j$ for the number of site, can be written as:
\begin{align}
\nonumber
\mathcal{S}_{ex}[c^\dagger,c;\boldsymbol{h}] &= 
\int_0^\beta d\tau \Biggl[ 
\sum_{k\sigma} c^\dagger_{k\sigma}(\tau) \bigl( \varepsilon_k + \partial_\tau \bigr) c_{k\sigma}(\tau) 
 \\ 
 \nonumber &+ U\sum_j \bigl(c_{j\uparrow}^\dagger(\tau)c_{j\uparrow}(\tau)-\tfrac12\bigr)
\bigl(c_{j\downarrow}^\dagger(\tau)c_{j\downarrow}(\tau)-\tfrac12\bigr)
\\ &- \sum_{j \sigma \sigma'} c_{j\sigma}^\dagger(\tau)\,\sigma^\alpha_{\sigma \sigma'}\,c_{j\sigma'}(\tau)\,h_j^\alpha \Biggr],
\end{align}
with dispersion law $\varepsilon_k = -2t \cos(k)$, Pauli matrices $\sigma^\alpha$ and inverse temperature $\beta=1/T$. For the sake of brevity, we denote the second term by $U\theta$ in what follows.

\begin{figure}[h]
\centering
\includegraphics[width=0.4\textwidth]{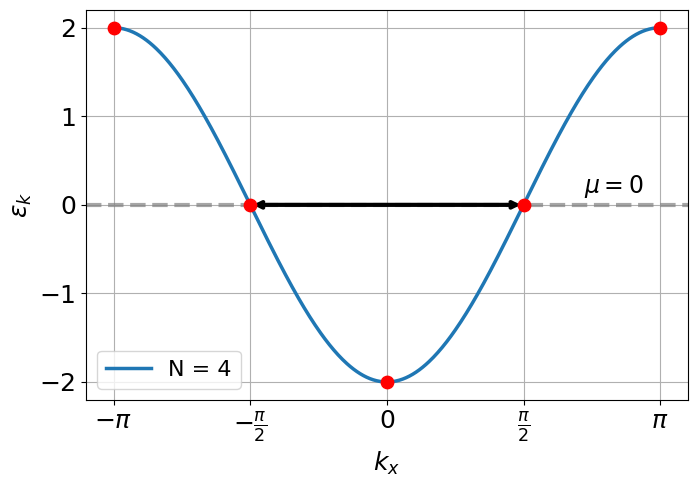} \\ \includegraphics[width=0.4\textwidth]{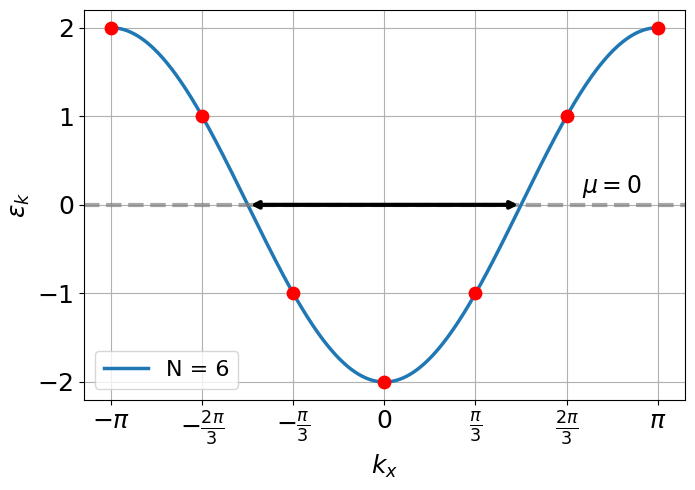} 
\caption{Dispersion law (solid line) and Fermi level (dashed line) of half-filled 1D Hubbard model with $N=4$ (upper panel) and $N=6$ (lower panel) sites. Occurrence of nesting between Fermi-points in $N=4$ case is observed.}
\label{fig:fig1}
\end{figure}

It is known that one-dimensional Hubbard chain possesses strong anti-ferromagnetic (AF) fluctuations at half-filling. The relevance indication can be seen from the Fermi surface (namely, points in 1D case) represented on Fig.~\ref{fig:fig1}. Perfect nesting between points $k=-\pi/2$ and $\pi/2$, which occurs at $K=\pi$, makes this system instable with respect to AF fluctuations. This observation as well as the fact, that low-dimensional systems are much more instable than 3D ones, makes this deceptively simple system a good toy model for investigation of fluctuations impact on its physics. 

The main idea of the FLF approach is to build an ensemble of so-called parental systems, each with fixed value of order parameter relevant for leading fluctuation channel, and each subjected to artificial fluctuating external field $\boldsymbol{\nu}$. For the system under consideration the relevant order parameter is AF lattice spin, defined as:
\begin{equation}
    s^\alpha = \frac{1}{\beta N} \sum_{j\sigma\sigma'} \int_0^\beta c^\dagger_{j\sigma}(\tau) \sigma^\alpha_{\sigma\sigma'} c_{j\sigma'}(\tau) e^{i\pi j} d\tau.
\end{equation}
The described FLF approach can be realized via the exact transformation of partition function, which in our case takes the form
\begin{align}
\nonumber
    \mathcal{Z} &= \int e^{-\mathcal{S}_{ex}[c^\dagger,c;\boldsymbol{h}]}\mathcal{D}[c^\dagger,c] \\
    &= \mathcal{C}\iint e^{-\mathcal{S}_{ex}[c^\dagger,c;\boldsymbol{h}]-\frac{\beta N}{2\lambda}(\boldsymbol{\nu} - \boldsymbol{h}-\lambda \mathcal{\boldsymbol{s}})^2}\mathcal{D}[c^\dagger,c]d^3\nu \\
    &= \mathcal{C}\iint e^{-\mathcal{S}_\nu^0[c^\dagger,c]+U\theta + \frac{\beta N \lambda}{2}s^l s^l}\mathcal{D}[c^\dagger,c]d^3\nu.
    \label{eq:FLF_transform}
\end{align}
The overall constant $\mathcal{C}= \Bigl(\frac{\beta N}{2\pi \lambda}\Bigr)^{3/2}$ will be canceled during calculation of any physical observable, so we will omit it in what follows. $\lambda$ is the free parameter of this transformation, whose value can be chosen so that the effective long-range interaction $\sim s^ls^l$ emerging in \eqref{eq:FLF_transform} efficiently depresses the true Hubbard interaction $U \theta$ \cite{ys24}. Under this choice, the resulting FLF partition function can be expressed as
\begin{equation}
    \mathcal{Z} \approx \mathcal{Z}_{\text{FLF}} = \int \mathcal{Z}_0(\boldsymbol{\nu}) e^ {-\frac{\beta N}{2\lambda}(\boldsymbol{\nu} - \boldsymbol{h})^2}  d^3 \nu,
    \label{eq:generalFLFU0}
\end{equation}
where
\begin{equation}
    \mathcal{Z}_0(\boldsymbol{\nu}) = \int e^{-\mathcal{S}_\nu^0[c^\dagger,c]}\mathcal{D}[c, c^\dagger]
    \label{eq:Z_nu_FLF0}
\end{equation}
is the partition function of non-interacting Hubbard chain subjected to external artificial AF field $\boldsymbol{\nu}$ with the action
\begin{align}
\nonumber
\mathcal{S}^0_\nu[c^\dagger,c]  = \int_0^\beta d\tau \Biggl( \sum_{k\sigma} c^\dagger_{k\sigma}(\tau) &\left( \varepsilon_k + \partial_\tau \right) c_{k\sigma}(\tau)  \\-
&\sum_{k\sigma\sigma'} c^\dagger_{k\sigma}(\tau)\sigma^\alpha_{\sigma\sigma'} c_{k+\pi,\sigma'}(\tau) \nu^\alpha \Biggr),
\label{eq:S_nu_classical}
\end{align}
which is now fully diagonalized in momentum space.

To calculate a physical observable $\hat{\mathcal{O}}$ in the FLF approach, we should, at first, take its average $\langle\mathcal{O}\rangle_{\boldsymbol{\nu}}$ over ensemble with partition function $\mathcal{Z}_0(\boldsymbol{\nu})$, and then average it over all values of fluctuating field $\boldsymbol{\nu}$, namely:
\begin{equation}
    \langle \mathcal{O}\rangle_{\text{FLF}} = \mathcal{Z}_{\text{FLF}} ^{-1}\int \langle\mathcal{O}\rangle_{\boldsymbol{\nu}}\mathcal{Z}_0(\boldsymbol{\nu}) e^{-\frac{\beta N}{2\lambda}(\boldsymbol{\nu} - \boldsymbol{h})^2} d^3\nu \label{eq:averagenonop}.
\end{equation}
In the absence of external field $\boldsymbol{h}$ and for isotropic quantities $\hat{\mathcal{O}}$ this integral can be reduced to one-dimensional:
\begin{equation}
    \langle \mathcal{O}\rangle_{\text{FLF}} = \frac{\int \langle\mathcal{O}\rangle_{\boldsymbol{\nu}}\mathcal{Z}_0(\boldsymbol{\nu}) e^{-\frac{\beta N}{2\lambda}\nu^2} \nu^2d\nu}{\int \mathcal{Z}_0(\nu) e^ {-\frac{\beta N}{2\lambda}\nu^2}  \nu^2 d\nu}
    \label{eq:neweq}
\end{equation}

\section{Accounting for quantum fluctuations in the Quantum FLF scheme}
\label{section:Q-FLF}
In this work we aim to investigate the impact of quantum fluctuations on the physics of correlated fermions using the FLF scheme as a research tool. We thus consider non-zero Matsubara frequencies of fluctuating field $\boldsymbol{\nu}$ and extend the basic FLF approach with the use of $\tau$-representation:
\begin{equation}
\label{eq:nu_tau_full}
   \nu^{\alpha}(\tau) =  \sum_{m=-\infty}^{+\infty} \nu_m^\alpha e^{iw_m\tau}.
\end{equation}
Here $\tau$ is the imaginary time, and $w_m=2 \pi m/\beta$ is the $m$th bosonic Matsubara frequency of $\alpha$-component of field $\boldsymbol{\nu}$. Rewriting also \eqref{eq:S_nu_classical} in Matsubara frequency representation, and substituting \eqref{eq:nu_tau_full} in the resulting expression we obtain:
\begin{equation}
\mathcal{S}^0_\nu[c^\dagger,c] = \sum_{\substack{nk\\\sigma}} c^\dagger_{k\sigma n}(\varepsilon_k - i\omega_n)c_{k\sigma n} - \sum_{\substack{nmk\\\sigma\sigma'}} c^\dagger_{k\sigma n}\sigma^\alpha_{\sigma\sigma'}c_{k+\pi,\sigma',n+m}\nu_m^\alpha,
\label{eq:General_action}
\end{equation}
where $\omega_n=(2n+1)\pi/\beta$ is the $n$th fermionic Matsubara frequency.

Restricting our investigation to intermediate temperatures, we limit the number of fluctuating field Matsubara modes to 3. Namely we will consider only the mode  $m=0$, and the very first non-zero modes $m=\pm 1$ (see \cite{ys24}, where the number of momentum modes was limited due to intermediate spatial size of considered system). Moreover, revealing, whether the quantum fluctuations influence the physical observables in general, it is reasonable to focus on modes $m=\pm 1$ as a small corrections to the impact of $m=0$. This reasoning is even stronger if we take into account that the use of the static field $\nu^\alpha$ in FLF scheme is already enough for eliminating unphysical N\'eel transition for low-dimensional systems \cite{rubtsov2020collective,ys24}. In what follows we use the subscript ``$0,\pm 1$'' for quantities calculated in this approximation.

Under these assumptions, we can generalize \eqref{eq:Z_nu_FLF0} and write a perturbation series for the logarithm of partition function, which up to the second order takes the form:
\begin{gather}
    \ln\left(\frac{\mathcal{Z}_{0,\pm 1}(\nu)}{\mathcal{Z}_\nu^{(0)}}\right) \approx 
    \frac{\mathcal{Z}_\nu^{(1)}}{\mathcal{Z}_\nu^{(0)}} + 
    \frac{\mathcal{Z}_\nu^{(2)}}{\mathcal{Z}_\nu^{(0)}} - 
    \frac{1}{2} \left( \frac{\mathcal{Z}_\nu^{(1)}}{\mathcal{Z}_\nu^{(0)}} \right)^2.
    \label{eq:lnZexpansion}
    \end{gather}
Here $\mathcal{Z}_\nu^{(0)}, \mathcal{Z}_\nu^{(1)}$ and $\mathcal{Z}_\nu^{(2)}$ are the relevant terms of the expansion:
\begin{gather}
    \mathcal{Z}_{0,\pm 1}(\boldsymbol{\nu}) \approx \int  e^{-\mathcal{S}^0_{\nu_0}}  \Bigl(1 -\mathcal{S}_{\nu_{\pm 1}} + \frac{1}{2}\mathcal{S}_{\nu_{\pm 1}}\mathcal{S}_{\nu_{\pm 1}} \Bigr)\mathcal{D}[c^\dagger,c],
    \label{eq:expansion}
\end{gather}
and $\mathcal{S}^0_{\nu_0}$ is the action of non-interacting system in external field $\boldsymbol{\nu}_0$, and $\mathcal{S}_{\nu_{\pm1}}$ corresponds to coupling between non-zero Matsubara modes $\boldsymbol{\nu}_{\pm 1}$ and AF-order parameter, and is considered as a perturbation. Full expressions for these actions are presented in Appendix \hyperref[app:partition]{A}. 

Since linear terms vanish because of symmetric Gaussian distribution of the fields $\boldsymbol{\nu}$, the expression for the Q-FLF partition function takes the form:
\begin{gather}
    \mathcal{Z}_{0,\pm 1}(\boldsymbol{\nu}) \approx \mathcal{Z}_{\nu}^{(0)}\exp \left(\frac{\mathcal{Z}_{\nu}^{(2)}}{\mathcal{Z}_{\nu}^{(0)}}\right).
\end{gather}

Now we are ready to construct the generalization of \eqref{eq:generalFLFU0} for the Q-FLF case:
\begin{equation}
\mathcal{Z}_{\text{Q-FLF}} = \int \mathcal{Z}_{0,\pm 1}(\boldsymbol{\nu})
e^{-\frac{\beta N}{2\lambda}\left[(\boldsymbol{\nu}_0 - \boldsymbol{h})^2 + |\boldsymbol{\nu}_1|^2 + |\boldsymbol{\nu}_{-1}|^2\right]} d^3\nu_0 d^3\nu_{1}  d^3\nu_{-1}.
\label{eq:Z01dall}
\end{equation}
At vanishing $\boldsymbol{h}$, it is possible to integrate over $\boldsymbol{\nu}_{\pm 1}$ analytically to obtain:
\begin{gather}
     \mathcal{Z}_{\text{Q-FLF}} = \mathcal{C}_{0,\pm 1} \int \frac{\det B}{\det A} \nu_0^2 \exp \left( -\frac{\beta N}{2\lambda}\nu_0^2\ \right)d\nu_0,
     \label{eq:Z01}
\end{gather}
where constant $\mathcal{C}_{0,\pm1}$ may be omitted while calculating observables. Here $\det B$ is in fact the partition function of non-interacting system embedded in field $\boldsymbol{\nu}_0$, with matrix $B$ being
\begin{equation}
B_{\substack{k\sigma n\\k'\sigma'n'}} = 
\Bigl[ (\varepsilon_k - i\omega_n)\,\delta_{kk'}\otimes \delta_{\sigma\sigma'} 
- \nu^\alpha_0 \,\delta_{k+\pi,\,k'}\otimes \sigma^\alpha_{\sigma\sigma'} \Bigr] \otimes \delta_{nn'},
\end{equation}
and $\det A$ comes from contribution of $\boldsymbol{\nu}_{\pm 1 }$ fields. Full expression of $A$ and the relevant derivations are presented in Appendix \hyperref[app:partition]{A}. Further, to distinguish the case, where $\boldsymbol{\nu}$ includes zeroth mode only, we will mark the relevant quantities with $\text{FLF}_0$ subscript.

Now, when the general scheme of Q-FLF is developed, we move on to calculation of Green's function, and a number of physical observables to investigate the symptoms of quantum fluctuation modes impact.

\section{Green's function}
\label{section:GF}
Green's function appears as a versatile instrument for analysis of different aspects of physical systems such as e.g. spectral function. It represents the spectrum of quasi-particle excitations, which, moreover, may be experimentally measured \cite{APRES}.

Since we are working with non-zero temperatures, Matsubara imaginary time formalism will be used. General expression for Green's function in this case reads:
\begin{equation}
\mathcal{G}_{kk'\sigma\sigma'}(\tau, \tau') = -\left\langle \hat{T}_\tau  c_{k\sigma}(\tau) c_{k' \sigma'}^\dagger(\tau') \right\rangle,
\end{equation}
where $\hat{T}_\tau$ is the time-ordering operator. Under the natural assumption $\mathcal{G}(\tau,\tau') =\mathcal{G}(\tau -\tau')$ we can rewrite it in frequency domain as
\begin{equation}
    \mathcal{G}_{kk'\sigma\sigma'}(i\omega_n) = - \left\langle c_{k\sigma n} c^{\dagger}_{k'\sigma' n}\right\rangle.
\end{equation}

In order to assess the contribution of including non-zero Matsubara frequencies and to demonstrate the derivation principle on less computationally complex case, we start with the calculation of $\text{FLF}_0$ Green's function.

\subsection{Zero-mode FLF Green's function}

Derivation of zero-mode Green's function expression comes down to the use of formula \eqref{eq:neweq}:
\begin{equation}
\mathcal{G}^{\mathrm{FLF}_0}_{kk'\sigma\sigma'}(i\omega_n) 
= \mathcal{Z}_{\mathrm{FLF}_0}^{-1} \int 
\bigl\langle c^{\dagger}_{k'\sigma' n}c_{k\sigma n}  \bigr\rangle_{\boldsymbol{\nu}_0}
\; e^{-\frac{\beta N}{2\lambda}\boldsymbol{\nu}_0^2} \, d^3\nu_0.
\label{average}
\end{equation}
The inner averaging over the non-interacting parental approximation reads:
\begin{align}
\langle c^{\dagger}_{k'\sigma' n}c_{k\sigma n}\rangle_{\boldsymbol{\nu}_0}=
\int c^{\dagger}_{k'\sigma' n} c_{k\sigma n}
\exp\left(-\sum_{\substack{qs l\\q's'nl}} c_{qs l}^\dagger B_{\substack{qs l\\q's'l'}} c_{q's'l'}\right)
\,\frac{\mathcal{D}[c^\dagger,c]}{\mathcal{Z}_0(\boldsymbol{\nu}_0)}.
\end{align}
Integrating over Grassmann number gives
the result for Green's function of the non-interacting system in external field $\boldsymbol{\nu_0}$:
\begin{equation}
\begin{split}
\mathcal{G}^{\textrm{FLF}_0}_{kk'\sigma\sigma'}(i\omega_n;\boldsymbol{\nu}_0) = 
-\frac{ (\varepsilon_{k+\pi} - i\omega_n)\,\delta_{kk'}\delta_{\sigma\sigma'}
+ \nu_0\,\sigma^z_{\sigma\sigma'}\,\delta_{k+\pi,k'} }
{(\varepsilon_k - i\omega_n)(\varepsilon_{k+\pi} - i\omega_n) - \nu_0^2}.
\label{eq:GFLF0}
\end{split}
\end{equation}
Now, all we have to do is to integrate the expression \eqref{eq:GFLF0} over Gaussian ensemble of external fields $\nu_0$ according to formula \eqref{eq:neweq}, which we do numerically.

\subsection{Quantum FLF Green's function}
Now we move to accounting for non-zero Matsubara modes of fluctuations in Green's function. This calculation is also based on expansion of partition function \eqref{eq:expansion}. The first order terms of $\mathcal{G}^{\text{Q-FLF}}_{kk'\sigma\sigma'}(i\omega_n;\boldsymbol{\nu})$ can be obtained directly by calculating integrals over Grassmann variables, which is done in Appendix \hyperref[app:Greens]{B}. In Appendix \hyperref[app:Diagrams]{C} we show, that higher order corrections to Green's functions vanishes fast, using Feynman diagrams technique. Numerical calculations of the second-order term shown, that relative contribution of such diagrams occurs to be comparatively small ($\sim 10^{-3}$ -- $10^{-4}$). So, we retain only first-order terms of Green's function in what follows.

\subsection{Results for Green's function}

\begin{figure}[t]
\begin{minipage}{\linewidth}
\centering
\includegraphics[width=\linewidth]{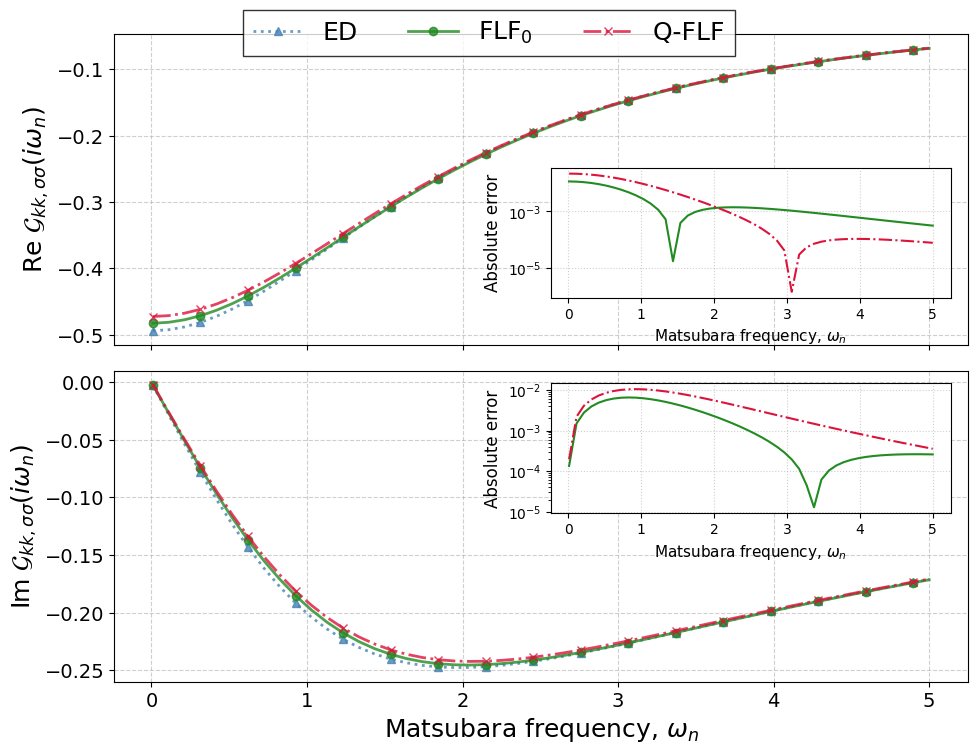} \\
\small\textbf{(a)}
\\
\includegraphics[width=\linewidth]{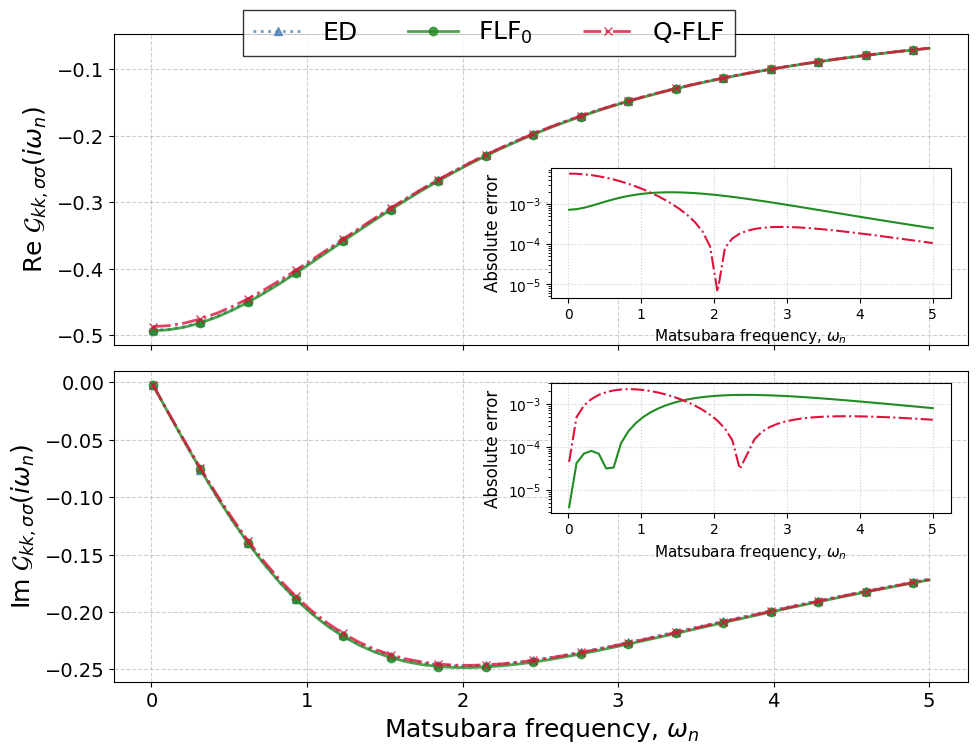} \\
\small\textbf{(b)}
\end{minipage}
\caption{Matsubara Green's function for Hubbard chain of $N=4$ (a) and $N=6$ (b) sites at $\beta=8$ for momentum $k=\pi$ in regime of moderate correlations $U/t=1$.}
\label{fig:GF_46}
\end{figure}

Results for the zero-mode and Q-FLF Green's function are shown on Figure \ref{fig:GF_46}, as well as reference results of exact diagonalization. At first, we see that $\text{FLF}_0$ approximation almost exactly reproduces reference data. But what is noteworthy, is that including of $\pm 1$ Matsubara modes into consideration in Q-FLF approach makes the results slightly worse. There are two possible reasonings for this effect. The first one concerns breaking the perturbation series at the second order in \eqref{eq:lnZexpansion}, which can be not enough to grasp the physics of excitations, and we need more orders to describe Green's function correctly. The second reason deals with the number of Matsubara modes we included in our calculations. Taking $\pm 1$ modes into account introduces additional excitation channel for fermions. It differs from the real picture though, as the system actually possesses all the Matsubara excitations spectrum interacting with each other. Frequency-dependent Green's function feels this artificial uncompensated $\pm 1$ excitation, which manifests in Q-FLF points standing further from exact ones. Returning to the observation that $\text{FLF}_0$ approach gives results close to exact for the Green's function, we come to the conclusion, that dynamical collective modes are not significant for this quantity, at least for the 1D Hubbard system under consideration. In the next two sections we will see, though, that they do impact the quantities, which absorb all fermionic frequencies. And here we start with the total energy.

\section{Total energy} 
\label{section:energy}
\begin{figure}[t]
\begin{minipage}{\linewidth}
\centering
\includegraphics[width=\linewidth]{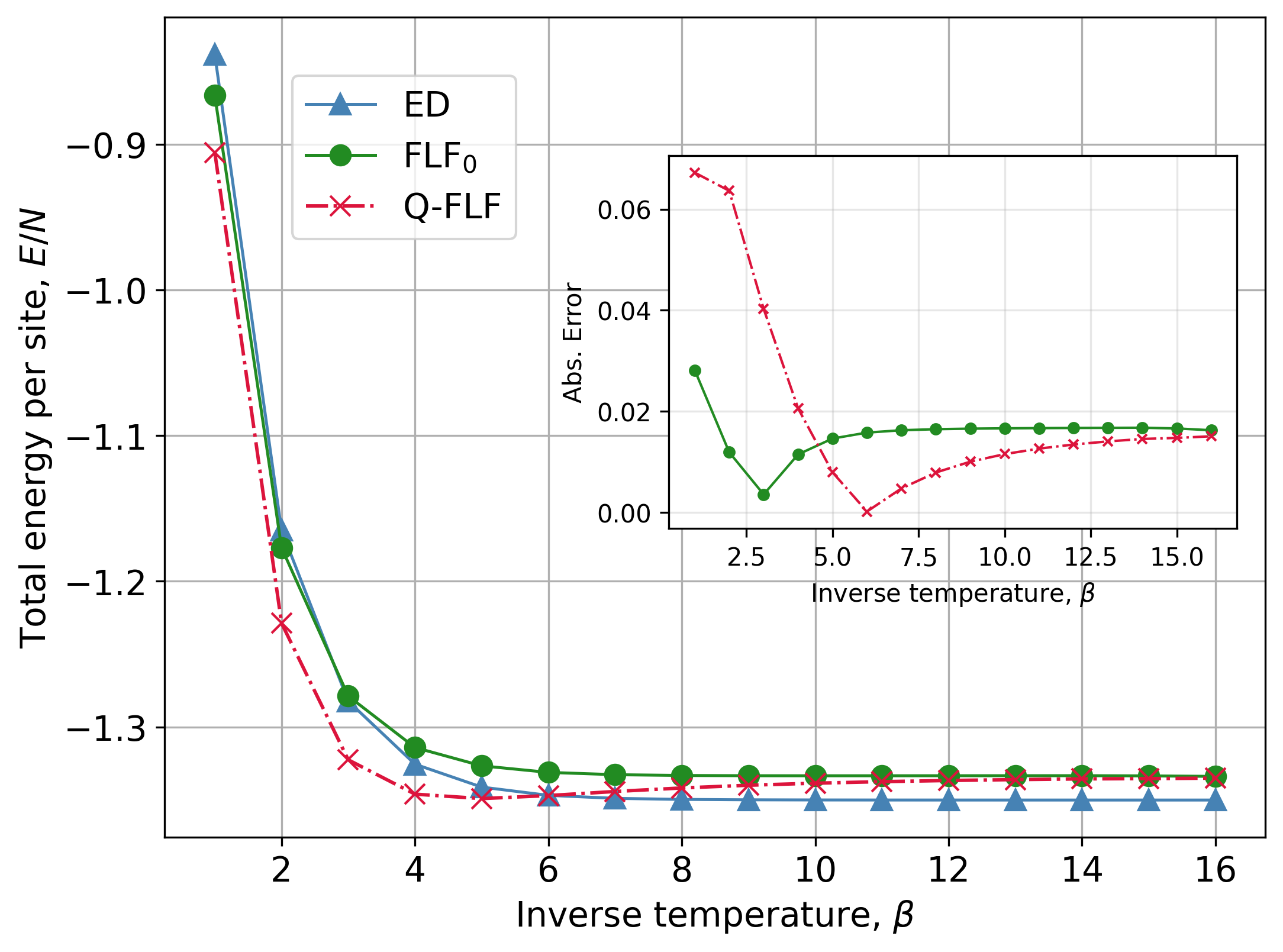} \\
\small\textbf{(a)}\\
\includegraphics[width=\linewidth]{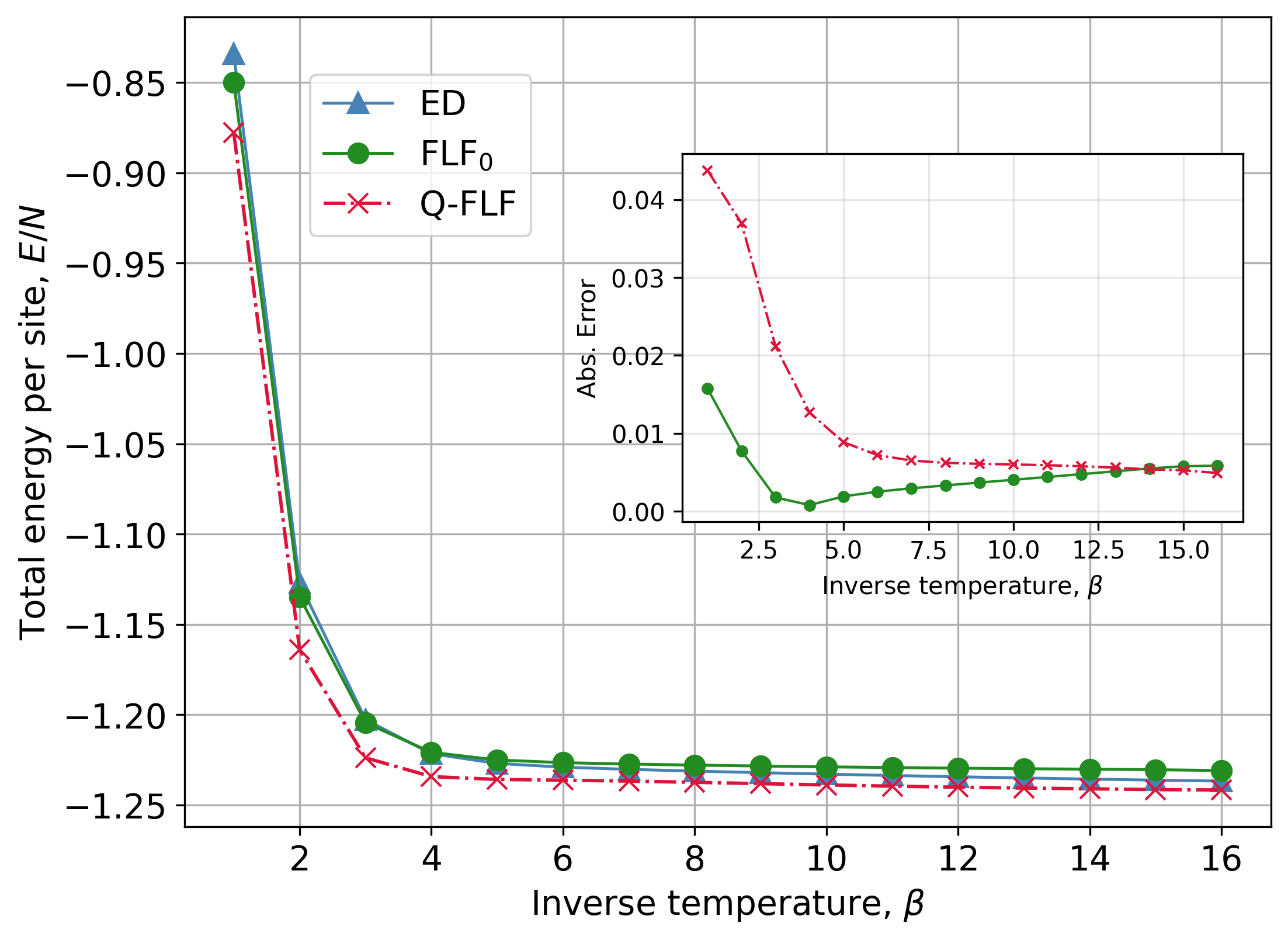} \\
\small\textbf{(b)}
\end{minipage}
\caption{Total energy per site $E/N$ as a function of
inverse temperature $\beta$ for 1D half-filled Hubbard chain with $N=6$ (a) and $N=8$ (b) sites at $U / t = 1$. The comparison is between the results obtained with Q-FLF, $\text{FLF}_{0}$ and exact diagonalization.}
\label{fig:energy}
\end{figure}

In this section we consider the total energy per site of the Hubbard chain by definition:
\begin{equation}
    \frac{E}{N} = -\frac{1}{N}\partial_\beta \ln \mathcal{Z}.
\end{equation}
We conduct the differentiation numerically, and compare $E/N$ in $\textrm{FLF}_0$ and Q-FLF approximations with the exact one, obtained via exact diagonalization (ED). The results are presented on Figure \ref{fig:energy}. From Fig. \ref{fig:energy}a one can see that inclusion of non-zero Matsubara modes allows to qualitatively improve results at intermediate temperatures, as we expected. It confirms that non-zero modes have at least quantitative impact on the physics of correlated systems. It is noteworthy that the mentioned improvement manifests for $N=6$ chain, but not for $N=8$ one. To make the reason clear, consider Fig. \ref{fig:fig1}. We see that the perfect nesting at AF vector $K=\pi$ occurs only for chains with $N$ which are multiples of $4$. AF fluctuations are well developed in such systems, and strongly influence its physics, so that the impact of non-zero modes is minor. On the other hand, this impact becomes pronounced for systems with other values of $N$.

\section{Susceptibility in the AF-channel}
\label{section:chi}
\begin{figure*}[t]
\centering

\begin{minipage}{0.47\textwidth}
\centering
\includegraphics[width=\linewidth]{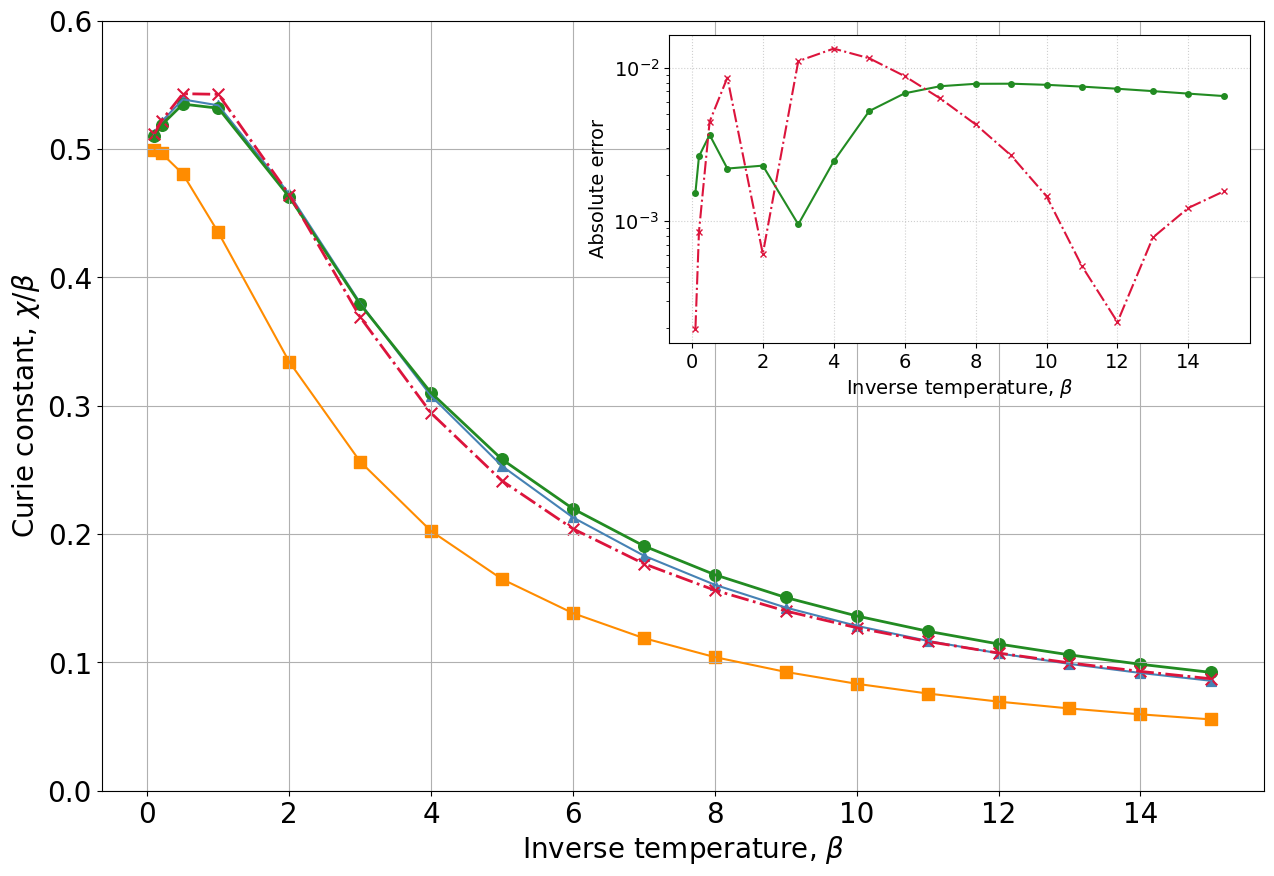}
\small\textbf{(a)}
\end{minipage}
\hfill
\begin{minipage}{0.47\textwidth}
\centering
\includegraphics[width=\linewidth]{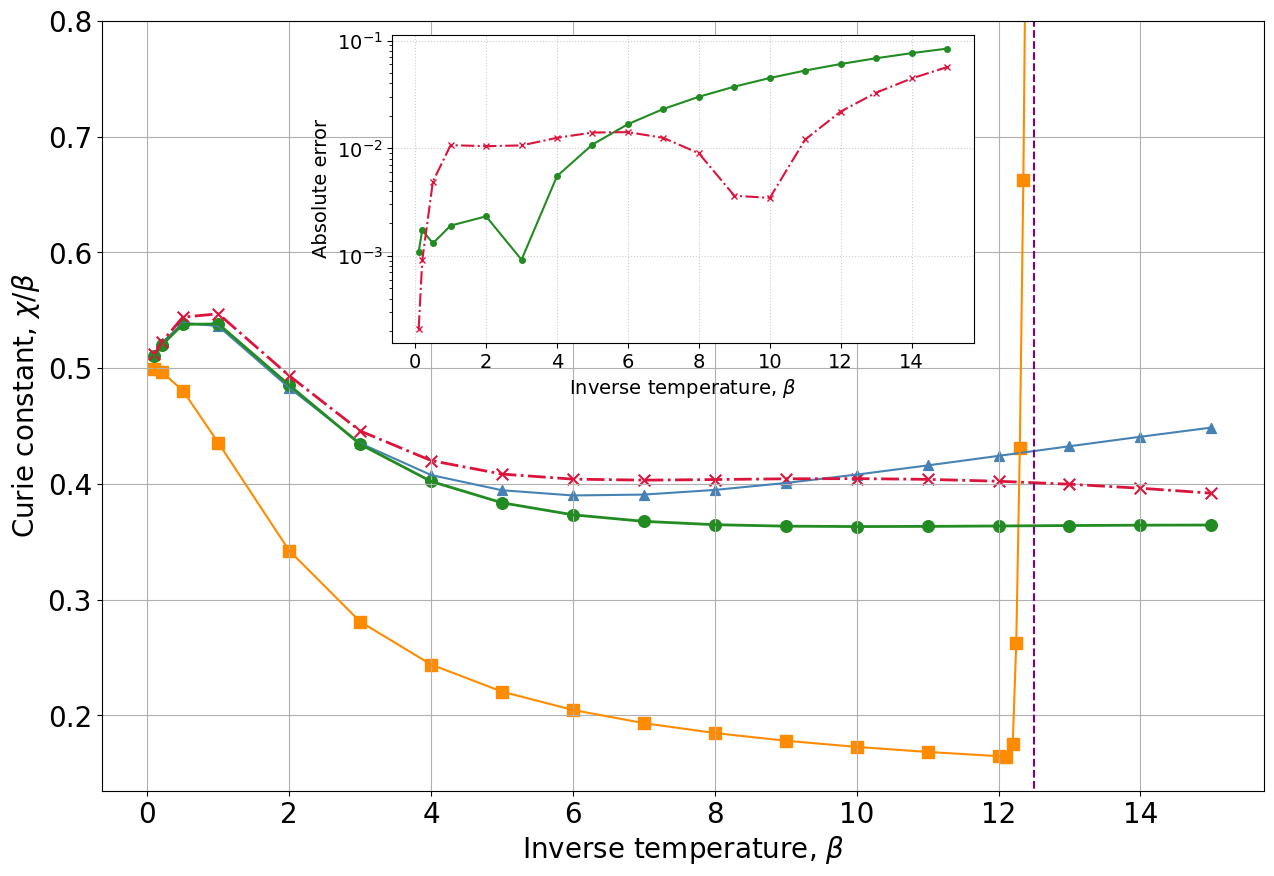}
\small\textbf{(b)}
\end{minipage}

\vspace{0.5cm}

\begin{minipage}{0.47\textwidth}
\centering
\includegraphics[width=\linewidth]{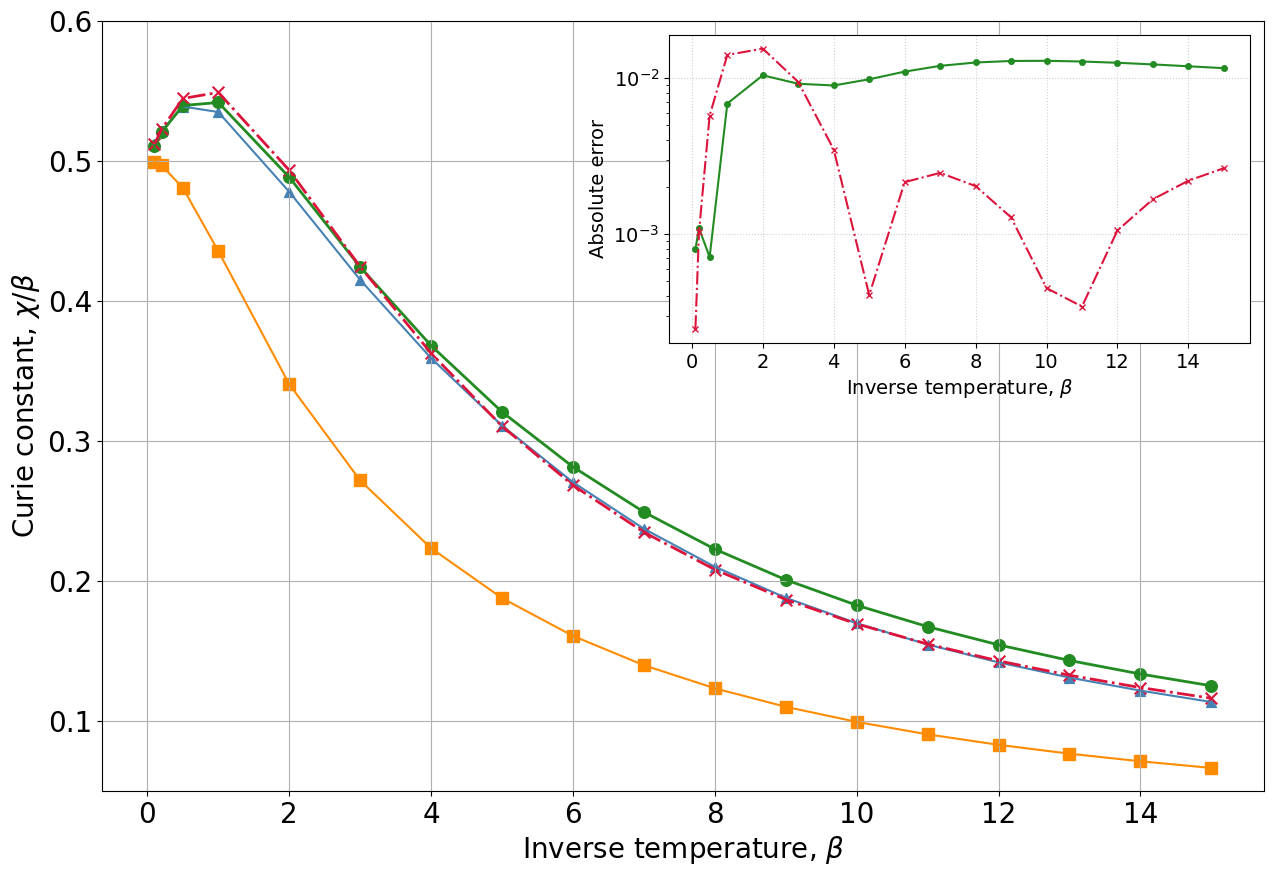}
\small\textbf{(c)}
\end{minipage}
\hfill
\begin{minipage}{0.47\textwidth}
\centering
\includegraphics[width=\linewidth]{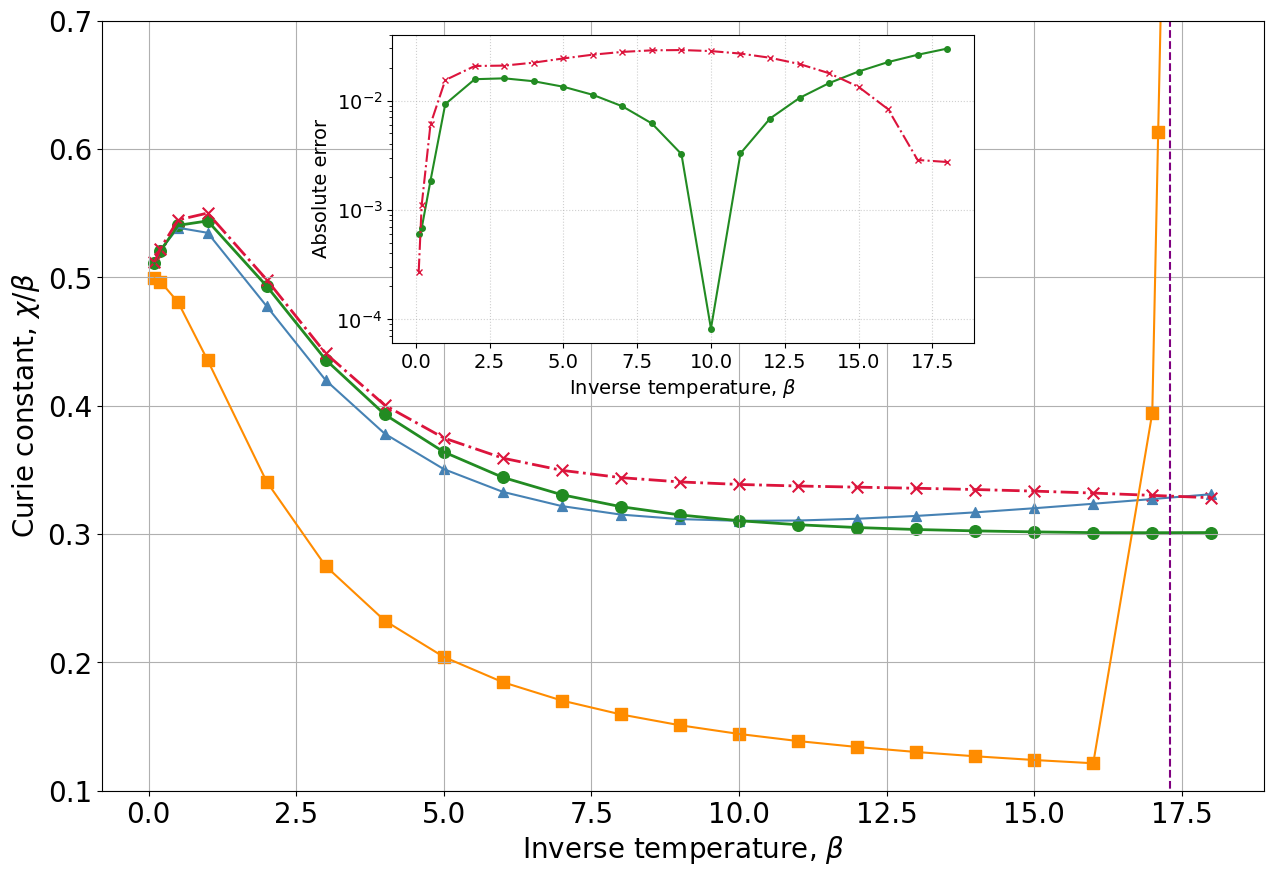}
\small\textbf{(d)}
\end{minipage}

\begin{minipage}{\textwidth}
\centering
\includegraphics[width=0.6\textwidth]{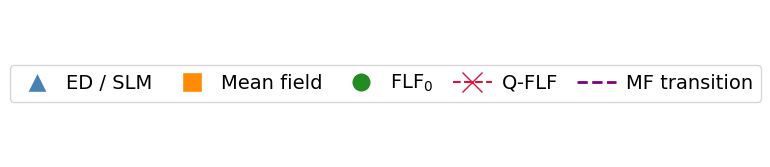}
\end{minipage}

\caption{Temperature dependence of Curie constant $\chi/\beta$ for the Hubbard chain of (a) N = 6, (b) N = 8, (c) N = 10 and (d) N = 12 in regime of moderate correlations $U/t = 1$ in different approximations.}
\label{fig:chi}
\end{figure*}

Susceptibility is often the tool of first choice for investigating response of the system to what is happening with its magnetic order parameter. In our case it is reasonable to talk about AF susceptibility, which means to consider susceptibility of the system to weak AF-ordered external field:
 \begin{equation}
\chi_{AF} = -\frac{1}{\beta N}\left.\frac{\partial^2 \ln \mathcal{Z}}{\partial h_{AF}^2} \right|_{h_{AF}\to 0}.
\end{equation}
Here we assumed that the tensor $\chi^{\alpha\gamma}_{AF}$ is isotropic, and denoted $\frac{1}{3}\sum_\alpha \chi^{\alpha\alpha}_{AF} \equiv \chi_{AF}$. Using expression (\ref{eq:generalFLFU0}) for $\textrm{FLF}_0$, and using \eqref{eq:Z01} for Q-FLF, we obtain:
\begin{equation}
    \chi_{AF}^{\textrm{FLF}_0/\textrm{Q-FLF}} = \frac{\beta N}{3\lambda^2}\langle\vec{\nu}_0^{\,2}\rangle_{\textrm{FLF}_0/\textrm{Q-FLF}} - \frac{1}{\lambda},
\end{equation}
depending on the approximation. Figure \ref{fig:chi} represents the results for the Curie constant $C(\beta) = \chi_{AF} / \beta$ of one-dimensional half-filled Hubbard chains with different number of sites. The $\text{FLF}_0$ and \text{Q-FLF} points were compared to the parental mean-field approximation (MF) and with results of numerically exact calculation, which was ED for $N\leq 8$ systems and stochastic Lanczos method (SLM) \cite{SLM} for $N>8$ systems.

On Figures \ref{fig:chi}b and \ref{fig:chi}d, we see that \text{Q-FLF} approach overcomes $\text{FLF}_0$ scheme for $N=8$ and $12$, which are multiples of $4$. The benefit from inclusion of non-zero Matsubara modes occurs at moderate and high values of $\beta$. It seems reasonable, as the lower the temperature -- the more significant is the role of quantum effects. It is interesting that the accuracy improvement achieved via inclusion of $\pm1$ Matsubara modes is more noticeable for $N=6$ and $N=10$ (Figures \ref{fig:chi}a and \ref{fig:chi}c). Less strong AF fluctuations at $\boldsymbol{\nu}_0$ mode, which is the case in the absence of perfect nesting, make Matsubara dynamics more significant on this background. Amplification of AF fluctuations for $N=8$ and $N=12$ also manifests as unphysical Ne\'el transition point marked by the MF AF susceptibility singularity.

\section{Discussion}
\label{section:discussion}
In this work, we investigated the possibility of selectively analyzing the contributions of dynamical collective fluctuations in correlated systems. As a computational framework, we developed an extended version of the Fluctuating Local Field (FLF) method that enables inclusion of non-zero Matsubara modes of collective fluctuations into consideration. This extension is referred to as the Quantum FLF (Q-FLF) approach. In principle, this scheme allows for the inclusion of an arbitrary number of non-zero Matsubara modes. But what is especially important for the purposes of this study, the Q-FLF method provides the capability to selectively switch individual modes on and off.

The role of non-zero Matsubara collective modes is examined for the one-dimensional half-filled Hubbard model with periodic boundary conditions, a system known to exhibit strong antiferromagnetic fluctuations. Our analysis is performed within a minimally non-static framework. Specifically, we restrict consideration to the contributions of the $\pm 1$ Matsubara modes in comparison to the zero-frequency mode. The proposed approach is applied to both single-particle and global properties of the system. In particular, we demonstrate that this minimal Q-FLF description is sufficient to quantitatively evaluate the contributions of the $\pm 1$ modes of antiferromagnetic fluctuations to the total energy and to the susceptibility in the antiferromagnetic channel. These contributions are found to be especially pronounced in the case where the number of sites in the one-dimensional Hubbard chain is not a multiple of $4$. In such systems, perfect nesting is absent, and consequently, the enhancement of antiferromagnetic fluctuations is reduced. As a result, the considered contributions are not masked by a strong antiferromagnetic background and therefore become more clearly observable.

An additional important result concerns the single-particle Green’s function. In this case, we observe a slight deterioration in accuracy upon inclusion of the $\pm 1$ modes compared to the zero-mode FLF approximation. We attribute this effect to the emergence of additional uncompensated fluctuation channels when only a limited subset of Matsubara modes is retained. In the exact system, such channels may interact nontrivially with other dynamical degrees of freedom. The resulting imbalance is particularly visible in sensitive quantities such as the Green’s function, which depends on fermionic Matsubara frequencies. In contrast, global observables such as the total energy and the antiferromagnetic susceptibility depend on the full spectrum of modes. For these quantities, the effect of such partial imbalance is significantly reduced. Consequently, even a minimal inclusion of non-zero Matsubara modes, such as the $\pm 1$ modes, yields an improvement over a static treatment. The results can also be interpreted in the context of the underlying methodology. The basic $\text{FLF}_0$ approach, and thus the Q-FLF scheme, is based on the assumption that the dominant contributions to the physical properties of the system arise from collective fluctuations, i.e., excitations involving a large number of degrees of freedom. It is therefore natural that the method provides a more accurate description of collective observables than of single-particle quantities. A more accurate treatment of the latter may require the inclusion of higher-order Matsubara modes.

We also note that improved results for the Green’s function can be obtained within perturbative frameworks that incorporate a more accurate treatment of the local Coulomb interaction. In the present work, the parameter $\lambda$ is chosen to fully compensate the on-site interaction. However, as shown in \cite{ys24}, this compensation is not exact. Thus, the present formulation effectively corresponds to the zeroth order of a perturbation expansion in the small parameter defined by the difference between the effective FLF interaction and the bare Hubbard interaction. Inclusion of higher-order corrections may further improve the results. While a direct comparison is not entirely straightforward, as the referenced study considers two-dimensional systems where fluctuations are weaker than in one dimension, extending this approach to the present context appears to be a promising direction for future research.

It is also important to emphasize that the contributions of individual Matsubara modes of collective fluctuations can be analyzed within other theoretical frameworks, including those discussed in the introduction, such as the dual boson approach \cite{RUBTSOV20121320} and related diagrammatic techniques. However, the Q-FLF method offers a significant computational advantage by mapping the original problem onto an ensemble of simpler systems. The computational procedure is thereby reduced to two main steps: evaluation of the partition function of a simple reference system (in our case non-interacting) and integration over auxiliary fluctuating fields with a Gaussian weight. This approach is typically less computationally demanding than the evaluation of diagrammatic expansions or dual boson schemes.

Finally, we stress that the restriction to the $\pm 1$ Matsubara modes in the present study is primarily motivated by the goal of testing the feasibility of a selective analysis of bosonic contributions. Moreover, as noted in the introduction, numerous studies indicate that key physical properties of correlated fermionic systems are governed by low-energy bosonic modes, particularly those associated with the zero and $\pm 1$ Matsubara frequencies. Our results are consistent with this expectation. At the same time, the present framework can be extended to include higher-frequency modes. In particular, it would be of interest in future work to investigate how higher-order Matsubara collective modes influence the Green’s function and spectral properties, and to establish a clear physical interpretation linking specific modes to underlying physical processes.

\section*{Acknowledgments}
The work of Ya.S. Lyakhova was supported by the Federal Academic Leadership Program Priority 2030 (NUST MISIS Strategic Technology Project “Quantum Internet”).

\appendix
\onecolumn
\section{Calculation of Q-FLF partition function}
\label{app:partition}
Considering $\mathcal{Z}_{0,\pm 1}(\boldsymbol{\nu})$ as the generalization of \eqref{eq:Z_nu_FLF0} and assuming the contribution of non-zero Matsubara modes $\boldsymbol{\nu}_{\pm 1}$ to be small, we construct perturbation series
\begin{gather}
    \mathcal{Z}_{0,\pm 1}(\boldsymbol{\nu}) \approx \int  e^{-\mathcal{S}^0_{\nu_0}}  \left(1 -\mathcal{S}_{\nu_{\pm 1}} + \frac{1}{2}\mathcal{S}_{\nu_{\pm 1}}\mathcal{S}_{\nu_{\pm 1}} \right)\mathcal{D}[c^\dagger,c], \label{expansion}
\end{gather}
where
\begin{equation}
\mathcal{S}^0_{\nu_{0}} =\sum_{\substack{kqn m\\\sigma\sigma'}} 
\left( c^\dagger_{k\sigma n}(\varepsilon_k - i\omega_n) c_{k\sigma' n} 
- c^\dagger_{k\sigma n} \sigma^\alpha_{\sigma\sigma'} \nu^\alpha_0 c_{k+\pi,\sigma' n} \right),
\label{eq:AppAS_0}
\end{equation}
\begin{equation}
\mathcal{S}_{\nu_{\pm 1}} = -\sum_{\substack{kqn m\\\sigma\sigma'}} 
\left( c^\dagger_{k\sigma n} \sigma^\alpha_{\sigma\sigma'} c_{k+\pi,\sigma', n+1} \nu^\alpha_1 
+ c^\dagger_{k\sigma n} \sigma^\alpha_{\sigma\sigma'} c_{k+\pi,\sigma', n-1} \nu^\alpha_{-1} \right).
\label{eq:AppAS_1}
\end{equation}
We can compute integrals over Grassmann variables analytically with the use of the following formula \cite{Altland}
\begin{equation}
    \int  c_i c_j c^\dagger_k c^\dagger_l \exp\left(-\sum_{s,j}c^\dagger_s B_{sj} c_j\right)\mathcal{D}[c^\dagger,c] =  \text{det}(B) \Bigl( B^{-1}_{il} B^{-1}_{jk} - B^{-1}_{ik} B^{-1}_{jl} \Bigl).
\end{equation}
In our case, matrix $B$ characterizes non-interacting system in external field $\boldsymbol{\nu}_0$
\begin{equation}
    B_{\substack{k\sigma n \\ k'\sigma'n'}} = (\varepsilon_k-i\omega_n) \delta_{kk'}\otimes \delta_{\sigma \sigma'}  \otimes \delta_{n n'} - \nu^\alpha_0 \delta_{k+\pi, k'}\otimes \sigma^\alpha_{\sigma \sigma'}  \otimes \delta_{n n'}.
\end{equation}

Computing these integrals and discarding the terms that vanish due to the structure of $B$-matrix, we see that
\begin{equation}
    \mathcal{Z}_{0,\pm1}(\boldsymbol{\nu}) \approx \det(B)\left(1 -\nu_1^\alpha \nu_{-1}^\gamma \sum_{
  \sigma\sigma'ss'
  \atop
  kqnm
} \sigma^\gamma_{ss'}\sigma^\alpha_{\sigma\sigma'}B^{-1}_{\substack{k+\pi,\sigma', n+1 \\ q,s,m}} B^{-1}_{\substack{q+\pi,s', m-1 \\ k,\sigma,n}}\right) \equiv \det(B) \left(1 - \nu_1^\alpha \nu_{-1}^\gamma  \Omega^{\alpha \gamma} \right).
\label{eq:Znucalculated}
\end{equation}
Matrix $\Omega$ includes summation over fermionic Matsubara frequencies, which can be calculated exactly by contour integration with following results:
\begin{gather}
    \Omega^{xx} = \Omega^{yy} = 4\beta\sum_q\frac{f(z_1)-f(z_2)}{E_q\left(E_q^2+\left(\frac{2\pi}{\beta}\right)^2\right)}\left(\left(\varepsilon_{q+\pi}-\varepsilon_q\right)^2+4\nu_0^2 \right)\\
    \Omega^{zz} =4\beta\sum_q\frac{f(z_1)-f(z_2)}{E_q\left(E_q^2+\left(\frac{2\pi}{\beta}\right)^2\right)}\left(\varepsilon_{q+\pi}-\varepsilon_q\right)^2,
\end{gather}
where $f(z) = \left(1 + e^{\,\beta z}\right)^{-1}$ is a Fermi function, $ z_1 = \sqrt{\varepsilon_q^2+\nu_0^2} = -z_2, \  E_q = 2z_1$ and  $\varepsilon_q = -2t\cos(q)$ is the dispersion. Non-diagonal elements of $\Omega$ are zero.

General expression for the Q-FLF partition function can be written as:
\begin{equation}
    \mathcal{Z}_{\text{Q-FLF}} = \int \mathcal{Z}_{0,\pm1}(\boldsymbol{\nu}) e^{-\frac{\beta N }{2 \lambda}\left(\boldsymbol{\nu}_0^2-|\boldsymbol{\nu}_1|^2 +|\boldsymbol{\nu}_{-1}|^2\right)}d^3\nu_0d^3\nu_1d^3\nu_{-1}.
\end{equation}
Using that $\boldsymbol{\nu}_{\pm1} =\boldsymbol{a} \pm i\boldsymbol{b}$ are complex conjugates one can decouple integrals over $\boldsymbol{\nu}_{\pm1}$ to obtain:
\begin{gather}
    \mathcal{Z}_{\text{Q-FLF}} = \det(B) \left\langle \left(\int\exp\left(-\frac{\beta N}{\lambda}\boldsymbol{a}^2\right)\exp\left(-\Omega^{\alpha \gamma}a^\alpha a^\gamma \right) d^3a\right)^2\right\rangle_{\text{FLF}_0},
\end{gather}
up to the overall constant. Integration over $\boldsymbol{a}$ can be done analytically with the use of well-known formula for Gaussian integrals:
\begin{equation}
    \int \exp\left( -a^\alpha M^{\alpha \gamma} a^\gamma \right)  d^3a = \sqrt{\frac{\pi^3}{\det M}}.
\end{equation}
It gives the final expression for $\mathcal{Z}_{\text{Q-FLF}}$ up to overall constant:
\begin{gather}
     \mathcal{Z}_{\text{Q-FLF}} = \int \frac{\det B}{\det A} \nu_0^2 \exp\left(-\frac{\beta N}{2\lambda}\nu_0^2\right)d\nu_0,
\end{gather}
where
$A^{\alpha \gamma} = \Omega^{\alpha\gamma} + \frac{\beta N}{\lambda}\delta^{\alpha\gamma}$.
 
\section{Q-FLF Green's function}
\label{app:Greens}
To calculate the Green's function, we follow the same procedure as for the partition function. Namely, we consider expansion
\begin{equation}
\mathcal{G}^{^{\textrm{Q-FLF}}}_{k\sigma n}(\boldsymbol{\nu})= \frac{1}{\mathcal{Z}_{0,\pm1}(\boldsymbol{\nu})
}\int c_{k\sigma n}c^\dagger_{k\sigma n} e^{-\mathcal{S}^0_{\nu_0}}\left(1 - \mathcal{S}_{\nu_{\pm 1}} + \frac{1}{2}\mathcal{S}_{\nu_{\pm 1}}\mathcal{S}_{\nu_{\pm 1}}\right)\mathcal{D}[c^\dagger,c].
\end{equation}
Substituting \eqref{eq:AppAS_0} and \eqref{eq:AppAS_1} here, and integrating over Grassmann variables, we obtain the following result for diagonal Green's function
\begin{align}
\mathcal{G}^{^{\textrm{Q-FLF}}}_{k\sigma n}(\boldsymbol{\nu}) 
&= B^{-1}_{\substack{k\sigma n\\k\sigma n}} 
+ \sum_{\substack{qq'mm'\\\eta\eta'\zeta\zeta'}}
\sigma_{\eta\eta'}^\alpha\nu_1^\alpha\sigma_{\zeta\zeta'}^\beta\nu_{-1}^\beta 
\Biggl[B^{-1}_{\substack{k\sigma n\\q\eta m'}}B^{-1}_{\substack{q'+\pi,\zeta',m-1\\k\sigma n}}B^{-1}_{\substack{q+\pi,\eta',m'+1\\q'\zeta m}} + B^{-1}_{\substack{k\sigma n\\q'\zeta m}}B^{-1}_{\substack{q+\pi,\eta',m'+1\\k\sigma n}}B^{-1}_{\substack{q'+\pi,\zeta',m-1\\q\eta m'}} \Biggr],
\end{align}

Here we calculate the diagonal component on momenta and spins. One can see, that non-diagonal momentum terms vanish due to expression of $B^{-1}$ (it will contain odd power of $\nu_0$, which vanish after integration over Gaussian ensemble). Remaining structure is equivalent for spin up and down. 
Further calculations involve substitution of the elements of inverse $B$ matrix in this expression and integration over ensemble of fluctuating fields $\boldsymbol{\nu}$ as in \eqref{eq:Z01dall}.

\section{Diagrammatic interpretation of Q-FLF approach}
\label{app:Diagrams}

It is instructive to develop diagrammatic approach for the Q-FLF. With this aim, we start with consideration of \eqref{eq:Znucalculated}. The first term in this expression stands for non-interacting case with static FLF field $\boldsymbol{\nu}_0$ only. The second term includes non-zero Matsubara modes of FLF field, and can be represented with a diagram 
\begin{equation}
\label{eq:linked}
    \begin{tikzpicture}[baseline=(current bounding box.center), scale=0.75, transform shape]
  \def\r{0.5}
  \coordinate (C) at (0,0);
  \draw[thick] (C) circle (\r);

  \coordinate (left)  at ($(C) + (180:\r)$);
  \coordinate (right) at ($(C) + (0:\r)$);

  \coordinate (Lext) at ($(C) + (180:1.2)$);
  \coordinate (Rext) at ($(C) + (0:1.2)$);

  \draw[decorate, decoration={snake, amplitude=0.3mm, segment length=3mm}, line width=0.5pt, -{Stealth[length=1.8mm]}]
    (Lext) -- (left);

  \draw[decorate, decoration={snake, amplitude=0.3mm, segment length=3mm}, line width=0.5pt, {Stealth[length=1.8mm]}-]
    (Rext) -- (right);

  \node at ($(Lext) + (-0.4,0)$) {$\nu_1^\alpha$};
  \node at ($(Rext) + (0.4,0)$) {$\nu_{-1}^\beta$};
\end{tikzpicture}
\end{equation}

Here straight line means zero-order Green's function $\mathcal{G}^{\textrm{FLF}_0}_{k\sigma}(i\omega_n;\boldsymbol{\nu}_0)$, and waves are generated by $\nu_1^\alpha\delta_{k',k+\pi} \otimes\sigma^\alpha_{\sigma \sigma'} \otimes \delta_{n',n-1}$ and $\nu_{-1}^\beta\delta_{k',k+\pi} \otimes\sigma^\beta_{\sigma \sigma'} \otimes \delta_{n',n+1}$. After integration over Gaussian ensemble of $\boldsymbol{\nu}_{\pm 1}$ fields the only non-zero terms occur to be of type $\langle \nu_1^\alpha \nu_{-1}^\beta \rangle_{\text{Q-FLF}}$. This way we obtain diagrams of kind:
\begin{equation}
\begin{tikzpicture}[baseline=(current bounding box.center), scale=0.75, transform shape]
\begin{feynman}
  \def\r{0.5}
  \coordinate (C) at (0,0);
  \draw[thick] (C) circle (\r);
  \coordinate (P0) at ($(C) + (0:\r)$);
  \coordinate (P180) at ($(C) + (180:\r)$);
  \draw[decorate, decoration={snake, amplitude=0.3mm, segment length=3mm}, line width=0.5pt]
    (P0) -- (P180);
\end{feynman}
\end{tikzpicture}
\hspace{1.5cm}
\begin{tikzpicture}[baseline=(current bounding box.center), scale=0.75, transform shape]
\begin{feynman}
  \def\r{0.4}
  \def\dist{1.6}
  \coordinate (L) at (0,0);
  \coordinate (R) at (\dist,0);
  \draw[thick] (L) circle (\r);
  \draw[thick] (R) circle (\r);
  \coordinate (Lup) at ($(L) + (30:\r)$);
  \coordinate (Rup) at ($(R) + (150:\r)$);
  \coordinate (Ldown) at ($(L) + (-30:\r)$);
  \coordinate (Rdown) at ($(R) + (-150:\r)$);
  \draw[decorate, decoration={snake, amplitude=0.3mm, segment length=3mm}, line width=0.5pt]
    (Lup) to[out=30, in=150, looseness=1] (Rup);
  \draw[decorate, decoration={snake, amplitude=0.3mm, segment length=3mm}, line width=0.5pt]
    (Ldown) to[out=-30, in=-150, looseness=1] (Rdown);
\end{feynman}
\end{tikzpicture}
\hspace{1.5cm}
\begin{tikzpicture}[baseline=(current bounding box.center), scale=0.75, transform shape]
\begin{feynman}
  \def\r{0.5}
  \def\dist{2.0}
  \def\yoffset{-0.25}
  \coordinate (C1) at (0,0);
  \coordinate (C2) at (\dist,\yoffset);
  \coordinate (C3) at (2*\dist,0);
  \draw[thick] (C1) circle (\r);
  \draw[thick] (C2) circle (\r);
  \draw[thick] (C3) circle (\r);
  \coordinate (L1) at ($(C1) + (0:\r)$);
  \coordinate (L2) at ($(C2) + (180:\r)$);
  \coordinate (M1) at ($(C2) + (0:\r)$);
  \coordinate (M2) at ($(C3) + (180:\r)$);
  \coordinate (LR1) at ($(C1) + (90:\r)$);
  \coordinate (LR2) at ($(C3) + (90:\r)$);
  \draw[decorate, decoration={snake, amplitude=0.3mm, segment length=3mm}, line width=0.5pt]
    (L1) -- (L2);
  \draw[decorate, decoration={snake, amplitude=0.3mm, segment length=3mm}, line width=0.5pt]
    (M1) -- (M2);
  \draw[decorate, decoration={snake, amplitude=0.3mm, segment length=3mm}, line width=0.5pt]
    (LR1) -- (LR2);
\end{feynman}
\end{tikzpicture}
 \, \, \dots
\end{equation}

This series contains all closed diagrams with wavy lines transmitting the unit bosonic Matsubara frequency and momentum $\pi$.

We can use this rule to construct diagrammatic representation of Green's functions. This way, the first order Q-FLF Green's function reads

\begin{equation}
\begin{tikzpicture}[baseline=(current bounding box.center)]

  \begin{feynman}
    \vertex (L1) at (0,0);
    \vertex (R1) at (1.2,0);
    \diagram* {
      (L1) -- [fermion, thick] (R1),
    };
  \end{feynman}
  \node at (1.8,0) {\(\approx\)};

  \begin{feynman}
    \vertex (L2) at (2.4,0);
    \vertex (R2) at (3.6,0);
    \draw[fermion thin] (L2) -- (R2);
  \end{feynman}
  \node at (4.2,0) {\(+\)};

  \begin{feynman}
    \vertex (L3) at (4.8,0);
    \vertex (R3) at (6.2,0);
    \vertex (v1) at (5.1,0);
    \vertex (v2) at (5.9,0);
    \draw[fermion thin] (L3) -- (v1);
    \draw[fermion thin] (v1) -- (v2);
    \draw[fermion thin] (v2) -- (R3);
    \draw[->, decorate, decoration={snake, amplitude=0.5mm, segment length=4mm}, line width=0.6pt]
      (v1) to[out=70, in=110, looseness=1] (v2);
  \end{feynman}
  \node at (6.8,0) {\(+\)};

  \begin{feynman}
    \vertex (L4) at (7.4,0);
    \vertex (R4) at (8.8,0);
    \vertex (w1) at (7.7,0);
    \vertex (w2) at (8.5,0);
    \draw[fermion thin]   (L4) -- (w1);
    \draw[fermion thin] (w1) -- (w2);
    \draw[fermion thin] (w2) -- (R4);
    \draw[<-, decorate, decoration={snake, amplitude=0.5mm, segment length=4mm}, line width=0.6pt]
      (w1) to[out=70, in=110, looseness=1] (w2);
  \end{feynman}

\end{tikzpicture}
\end{equation}

We can move on and consider the second order diagrams for the Green's function:
\begin{equation}
\begin{tikzpicture}[baseline=(current bounding box.center),
  fermionline/.style={draw=black, line width=0.5pt, postaction={decorate, decoration={markings, mark=at position 0.5 with {\arrow[scale=0.9]{stealth}}}}}
]

  \begin{feynman}
    \vertex (a0) at (0,0);
    \vertex (a1) at (0.7,0);
    \vertex (a2) at (1.4,0);
    \vertex (a3) at (2.1,0);
    \vertex (a4) at (2.8,0);
    \vertex (a5) at (3.5,0);
    
    \draw[fermionline] (a0) -- (a1);
    \draw[fermionline] (a1) -- (a2);
    \draw[fermionline] (a2) -- (a3);
    \draw[fermionline] (a3) -- (a4);
    \draw[fermionline] (a4) -- (a5);
    
    \draw[decorate, decoration={snake, amplitude=0.5mm, segment length=4mm}, line width=0.6pt]
      (a1) to[out=75, in=105, looseness=1.6] (a2);
    \draw[decorate, decoration={snake, amplitude=0.5mm, segment length=4mm}, line width=0.6pt]
      (a3) to[out=75, in=105, looseness=1.6] (a4);
  \end{feynman}

  \node at (4.2,0.2) {\large \(+\)};

  \begin{feynman}
    \vertex (b0) at (4.9,0);
    \vertex (b1) at (5.6,0);
    \vertex (b2) at (6.3,0);
    \vertex (b3) at (7.0,0);
    \vertex (b4) at (7.7,0);
    \vertex (b5) at (8.4,0);
    
    \draw[fermionline] (b0) -- (b1);
    \draw[fermionline] (b1) -- (b2);
    \draw[fermionline] (b2) -- (b3);
    \draw[fermionline] (b3) -- (b4);
    \draw[fermionline] (b4) -- (b5);
    
    \draw[decorate, decoration={snake, amplitude=0.5mm, segment length=4mm}, line width=0.6pt]
      (b1) to[out=75, in=105, looseness=1.3] (b3);
    \draw[decorate, decoration={snake, amplitude=0.5mm, segment length=4mm}, line width=0.6pt]
      (b2) to[out=75, in=105, looseness=1.3] (b4);
  \end{feynman}

\end{tikzpicture}
\end{equation}
where we assume 8 diagrams with each one corresponding to unique choice of wavy lines directions. Higher order diagrams can be constructed in the same way.

\twocolumn

\end{document}